\documentclass[oneside,12pt]{article}
{\topskip 2cm}
\usepackage{latexsym}
\usepackage{amsfonts}
\usepackage[dvips]{graphicx}
\begin{document}

\title{Vortex in Maxwell-Chern-Simons models coupled to external backgrounds}
\author{\large F. Chandelier$^a$, Y. Georgelin$^a$, M. Lassaut$^a$,\\ T.
Masson$^b$, J.C. Wallet
$^a$}
\date{}
\maketitle
\begin{center}
$^a$ Groupe de Physique Th\'{e}orique, 
Institut de Physique Nucl\'{e}aire, 
F-91406 Orsay CEDEX, France\\ 
$^b$ Laboratoire de Physique Th\'{e}orique, 
B\^{a}timent 210, 
Universit\'{e} de Paris XI, 
91405 Orsay CEDEX, France
\end{center}
\vskip 1 true cm

{\bf{Abstract}}: We consider Maxwell-Chern-Simons  
models involving different non-minimal coupling terms
to a non relativistic massive scalar and further coupled to an external
uniform background charge. We study how these models can be
constrained to support static radially symmetric vortex configurations
saturating the lower bound for the energy. Models involving Zeeman-type
coupling support such vortices
provided the potential has a "symmetry breaking" form and a relation between
parameters holds. In models where minimal coupling is supplemented by
magnetic and electric field dependant coupling terms, non trivial vortex configurations minimizing the energy
occur only when a non linear potential is introduced. The corresponding vortices are studied numerically.\par
\strut\thispagestyle{empty}

\vfill LPT-Orsay 04-33
\pagebreak
\setcounter{page}{1}

\section{Introduction}
Vortex solutions in (2+1)-dimensional field theories have received a
constant attention (for
reviews, see e.g. \cite{VORTREV}, \cite{ANREV}) motivated in part by the possible role played by vortices
in various phenomena of condensed matter physics, such as (Hight-$T_c$)
superconductors, Josephson junctions arrays or Quantum Hall Effect. Various field theory models have been considered,
starting from the Abelian-Higgs model \cite{BOG} and then extending to Chern-Simons
or Maxwell-Chern-Simons (MCS) \cite{MCS} theories coupled with relativistic or non
relativistic matter systems \cite{VORTCS}, \cite{JPI},
 \cite{VORTKOR}, \cite{VORTMCSNR},
\cite{BARAC}, \cite{HORVAT}, \cite{VORTMCSREL} together with (N=2)
supersymmetric extensions and/or non-abelian generalizations \cite{SUS},
\cite{VORTREV}.
Basically, these models have been shown to support finite energy vortices
which, in most cases, for a suitable choice for the matter potential,
saturate the lower bound of the energy for the considered system.\par
In this note, we consider two different types of MCS models coupled to a non
relativistic massive scalar, hereafter called type I and type II MCS models. 
Both are further coupled to an external
uniform background charge (a situation which may be of interest in condensed
matter systems). We study how each type of models can be
constrained to support static radially symmetric vortex configurations
saturating the lower bound for the energy. In type I models, the usual
minimal coupling of the MCS gauge potential to the scalar is supplemented by a magnetic
field dependant (Zeeman-type) coupling. Type II models involve the
non-minimal coupling introduced and discussed in \cite{JSGW} (see also
\cite{MCSK} and first of \cite{VORTMCSREL})
in which both magnetic and electric field dependant coupling terms appear.
The introduction within MCS theories with
matter of such a non-minimal coupling whose strength must be fixed to a specific
value has been proposed as a possible alternative way to describe (non
standard) composite anyonic objects and/or statistical transmutation. Basically, in this
approach, the statistical properties of the described anyonic (composite)
objects are controlled by a MCS gauge potential with suitable Pauli-type
coupling with matter \cite{JSGW}. In this spirit, type II models may be viewed as a
modification of the Landau-Ginzburg effective theory proposed in \cite{KLZ} to
describe some global physical features of the Quantum Hall Effect. In
this latter effective theory \cite{KLZ}, the statistical transmutation (between fermions and
bosons) is simply controlled by a Chern-Simons statistical gauge potential
minimally coupled with matter. This point is presented in the appendix.\par
We find that type I models support (static) vortices saturating the lower
bound for the energy provided the matter potential has a usual "symmetry
breaking" form (with minimum linked with the magnitude of the uniform
background) and a relation between the masses and the strength of the Zeeman
coupling holds. This is shown in section 2. In section 3, we consider type
II models in which the strenght of the non minimal coupling is fixed to the specific
value mentionned above. We show that these models do not support vortices saturating
the lower bound for the energy unless a non linear potential is introduced.
The corresponding vortex configurations are studied numerically. The case
where type II models are coupled with an external gauge potential is also
considered and gives rise to similar conclusions. In section 4, we summarize
the results and we conclude.\par

\vskip 0,3 true cm
\section{Vortex solutions in type I MCS models}
\vskip 0,3 true cm

The gauge invariant action for the first class of models we consider,
hereafter called type I models, is defined by{\footnote{Our conventions are
$\hbar=c=1, g_{\mu \nu}={\rm diag}(+,-,-),x=(x_0;{\bf x}=(x_1,x_2)).$
Polar coordinates are $x_1=r \cos \theta,x_2=r \sin \theta $. We define
${\bf e_{\theta}}=(\sin \theta,-\cos \theta )$}}
$$S_1=S_{MCS}+S_1^m; \ S_{MCS}=\int d^3x \big( -\frac{1}{4e^2} F_{\mu \nu} F^{\mu \nu}
 + \frac{\eta}{4} \epsilon_{\mu \nu \rho} A^{\mu} F^{\nu \rho}\big)
 \eqno(2.1a;b),$$ 
$$S_1^m=\int d^3x \big(i \phi^\dag{{D}}_0 \phi - \frac{1}{2 m} \vert {\bf{D}} \phi \vert^2 
+\kappa F_0
 \phi^\dag \phi -
 V(\phi) - A_0 J_0 \big) 
\eqno{(2.1c).}
$$
In (2.1), $F_{\mu \nu}=\partial_{\mu} A_{\nu}-\partial_{\nu} A_{\mu}, F_{\mu}
=\frac{1}{2} \epsilon_{\mu \nu \rho} F^{\nu \rho}$ is the dual field strength
whose time-like (resp. space-like) component $F_0$ (resp. $F_i, i=1,2$) is
associated to the magnetic (resp. electric) field,
$ \eta$ is the Chern-Simons coefficient,
$\phi$ is a non relativistic scalar with mass $m$, $
D_{\mu}=\partial_{\mu}-i A_{\mu}$ is the covariant derivative. 
$V(\phi)$ is the
potential to be specified in a while, $J_0$ denotes an external static
uniform background charge assumed to be positive and the 
term involving $\kappa$ is a Zeeman term. The mass dimensions for
the parameters and fields are
$[A_{\mu}]=[\phi]=1,[J_0]=2,[e^2]=[m]=1,[\eta]=0,[\kappa]=-1$. The equations of motion 
stemming from (2.1) are
$$
-\frac{1}{e^2} \ \epsilon_{i j} \partial^i F^j + \eta F_0 + j_0 - J_0 =0
 \eqno{(2.2a),}
$$
$$
\frac{1}{e^2} (\partial_i F_0 - \partial_0 F_i) + \eta \ \epsilon_{i j} F^j
+ \epsilon_{i j} j^j -\kappa  \ \partial^i j_0 =0 \,\  (i=1,2)
\eqno{(2.2b),}
$$
$$
iD_0 \phi + \kappa \ F_0 \phi + 
\frac{1}{2 m} \ D_iD_i \phi = 
 \frac{\delta V}{\delta\phi^\dag} \qquad\quad  ({\rm and \  h.c.} ) \ ,
\eqno{(2.2c),}
$$
where summation over space indices $i,j$ is understood ($\epsilon_{12}=+1$) 
and the components
of the gauge invariant matter current are given by
$$
j_0=\phi^\dag \phi ; \ \ j_i =\frac{i}{2m} (\phi^\dag D_i \phi-
(D_i \phi)^\dag\phi )
\eqno{(2.3a;b).}
$$
We now look for static radially symmetric solutions of the equations of
motion having finite energy and satisfying
$$
{\bf A} = {\bf e_{\theta}} \frac{a(r)+n}{r}, n \in \mathbb{Z};
 \ A_0(r)=a_0(r)
; \phi=f(r) e^{-i n \theta}
\eqno{(2.4a;b;c)}
$$
$$
\lim_{r \to \infty} a(r)=\lim_{r \to \infty} a_0(r)=0 ; \lim_{r \to 0} a(r)=-n
\eqno{(2.4d;e)}
$$
where $a(r)$ and $a_0(r)$ appearing in the usual vortex Ansatz (2.4a-c)
are smooth radial functions and (2.4e) corresponds to configurations
carrying a quantized magnetic flux $\Phi, \Phi=\int d^2x F_0= -2 \pi n $ where $n \in
\mathbb{Z}$ is related to the vorticity. The boundary conditions for $f$
will be determined in the course of the discussion.\par
>From now on, we drop the explicit radial 
dependence on the various functions to simplify the notations. Furthermore we define $X'=\frac{dX}{dr}$ for any
radial function $X$. Then, eqn.(2.2) can be conveniently reexpressed as
$$
-\frac{1}{e^2} \Delta a_0 -\eta \frac{a'}{r} + f^2 - J_0=0 \ ; 
 -\frac{1}{e^2} \left( \frac{a'}{r} \right)'-\eta a'_0 + \frac{f^2}{m}
 \frac{a}{r} -
 \kappa (f^2)' =0
\eqno{(2.5;a,b),}
$$
$$
\frac{1}{2m} \Delta f + f \left[ a_0  -\kappa\frac{a'}{r} \ -\frac{1}{2m}
\left(\frac{a}{r} \right)^2 \right]=\frac{\delta V}{\delta\phi^\dag}\vert_{\phi=f}
\eqno{(2.5c)}
$$
in which $\Delta=\frac{1}{r} \frac{d}{dr} (r \frac{d}{dr} .)$ The
Hamiltonian for the system (2.1) is given by
$${\cal{H}}_1=\int d{\bf{x}}\big({{1}\over{2e^2}}(F_0^2+{\bf{F}}^2)+{{1}\over{2m}}\vert{\bf{D}}\phi\vert^2-
\kappa F_0\phi^\dag\phi+V(\phi)\big) \eqno(2.6)$$ from which one obtains the
static energy functional density expressed in terms of the radial variables
(2.4) as
$$H_1=\int
d{\bf{x}}\big({{1}\over{2e^2}}[({{a^\prime}\over{r}})^2+a_0^{\prime2}]+{{1}\over{2m}}
[f^{\prime2}+f^2({{a}\over{r}})^2]+\kappa f^2{{a^\prime}\over{r}}+V(f)\big)
\eqno(2.7).$$
In the absence of any further requirements, type I models (2.1) involve 5 free
parameters. Furthermore, the potential $V(\phi)$ is
still unspecified. This later will have to be choosen in such a way
that the energy is definite positive, as it will be shown in the
sequel. We are now in position to select some specific models which admit finite energy vortex
solutions for some suitable choice for the potential. Notice that we will
focalize on vortex solutions saturating the minimum for the energy.\par 
To do this, one observes that the static energy (2.7) can be conveniently
rewritten as $H_1=H_1^\prime+H_{1V}$$+H_{10}$ with
$$H_1^\prime=\int d{\bf{x}}\big({{1}\over{2e^2}}(a_0^{\prime2}+[{{a^\prime}\over{r}}+
e^2\lambda_\pm(f^2-J_0)]^2)+
{{1}\over{2m}}[f^\prime\pm
f{{a}\over{r}}]^2\big) \eqno(2.8a)$$
$$H_{1V}=\int d{\bf{x}}
\big(V(f)-e^2{{\lambda_\pm^2}\over{2}}(f^2-J_0)^2\big)
\eqno(2.8b),$$
$$ H_{10}=\lambda_\pm J_0\Phi\mp[f^2a]_0^\infty \eqno(2.8c),$$
where $\lambda_\pm\equiv\kappa\pm{{1}\over{2m}}$, $[X]_0^\infty\equiv
X(\infty)-X(0)$ and we have explicitely collected the boundary terms in
$H_{10}$. Then, it can be readily observed that (2.8) has a
lower bound $H_{10}$ provided the potential is choosen to be
$$V(\phi)=e^2{{\lambda^2_\pm}\over{2}}(\phi^\dag\phi-J_0)^2 \eqno(2.9)$$ 
and that $H_{10}$ is saturated by field configurations satisfying
$$a_0=0;\ \ {{a^\prime}\over{r}}+e^2\lambda_\pm(f^2-J_0)=0;\ f^\prime\pm
f{{a}\over{r}}=0  \eqno(2.10a;b;c).$$
Consistency with the equations of motion then requires that
$$(1+e^2\eta\lambda_\pm)(f^2-J_0)=0; \ \Delta f-{{f^{\prime2}}\over{f}}=\pm 
e^2\lambda_\pm f(f^2-J_0) \eqno(2.11a;b).$$
Now one observes that (2.11a) is satisfied for any $f$ provided {\footnote{ Note that (2.11) admits the solution
$f^2=J_0$ which corresponds to the minimum for the potential (2.9).}}
$$\lambda_\pm=-{{1}\over{\eta e^2}} \eqno(2.12).$$
Then one
concludes that when (2.12) is satisfied, type I models with $V(\phi)$ given
by (2.9) have vortex
solutions of the form (2.4) whose matter density obeys
$$\Delta f-{{f^{\prime2}}\over{f}}=\mp{{1}\over{\eta}}f(f^2-J_0)
\eqno(2.13).$$ 
Moreover, these vortex configurations saturate the lower bound for the
energy. Notice that since $a_0=0$, these configurations have no electric
field. \par
Equation (2.13) has already appeared in various context (see e.g.
\cite{VORTREV}, \cite{BARAC}]). It is known to have
solutions whose asymptotic behavior is $f^2\sim J_0$ 
for $r\to\infty$
while for $r\to 0$, $f^2\sim r^p$ with $p\ge2$. Using these asymptotics together 
with (2.4d;e), one finds that
the second term in (2.8c) vanishes and $H_{10}$ reduces to
$H_{10}={{J_0}\over{e^2\eta}}\Phi$. The lower bound for the energy is
$\vert H_{10}\vert$ (within our conventions, positive (resp. negative) magnetic
flux corresponds to $\eta>0$ (resp. $\eta<0$) and to the lower (resp. upper)
sign in the above equations) which is proportional to the magnetic flux.
Integration over space of (2.5a) yields $\int d{\bf{x}}(f^2-J_0)$$+\eta\int
d{\bf{x}} F_0$$\equiv q+\eta\Phi=0$ so that the vortex configurations carry
a quantized charge proportional to the flux. \par
In the limit $\kappa=0$, (2.12) becomes
$\pm{{1}\over{2m}}=$$-{{1}\over{e^2\eta}}$. Selecting for instance $\eta>0$,
one concludes that in the absence of Zeeman term, the resulting type I model
still admits static vortex solutions saturating the lower bound for the
energy provided $m=M/2$ where $M$ is the mass
for $A_\mu$ (i.e the inverse screening lenght of the MCS $A_\mu$-mediated
interaction). Notice that this latter relation agrees with the one obtained
in \cite{BARAC} within MCS theory coupled with a massive non relativistic scalar in
the absence of Zeeman interaction term. \par
When $J_0=0$, still assuming that
(2.12) holds, (2.13) reduces to a simple Liouville equation while $H_{10}$
vanishes which is the situation considered in \cite{VORTMCSNR}. Again, selecting for
instance $\eta>0$, the corresponding solution for $f$ is given by
$f(r)={{2(n+1)}\over{r\eta^{1/2}}}(({{r}\over{r_0}})^{n+1}+$
$({{r_0}\over{r}})^{n+1})^{-1}$ where $r_0$ is some real constant. The
corresponding configuration carries a quantized electric charge Q since one
has from (2.5a) $\eta\Phi+$$\int d{\bf{x}}f^2$$\equiv\eta\Phi+Q=0$ with
$\Phi=-2\pi n$. Notice that one has now $f\to0$ for $r\to\infty$.\par
\vskip 0,3 true cm

\section{Vortex in type II MCS models}
\vskip 0,3 true cm
The present analyzis can be extended to another class of models, hereafter
called type II models, in which the coupling of the MCS sector to the non
relativistic matter involves terms depending on the electric field, in
addition to the magnetic coupling. We will first consider type II models
in the presence of an external background charge $J_0$ and then deal with
the coupling to an external gauge potential ${\cal{A}}_\mu$. In the first
case, the corresponding action is now defined by
$$S_2=S_{MCS}+S_2^m;\ S_2^m=\int
d^3x\big(i\phi^\dag{\cal{D}}_0\phi-{{1}\over{2m}}\vert{\cal{D}}_i\phi\vert^2-
V(\phi)-A_0J_0
\big)\eqno(3.1a;b)$$
where the operator ${\cal{D}}$ is given by
$$ {\cal D}_{\mu}=\partial_{\mu} - i A_{\mu}-i \kappa F_{\mu} 
\eqno{(3.2)}
$$
and $J_0$ is assumed to be positive as in section 2.
Time-like component of the term involving $\kappa$ gives rise to a magnetic
dipole (Zeeman)
coupling which is already present in $S_1^m$ (2.1b) while the corresponding 
space-like components generate gauge-invariant couplings depending on the
electric field, as announced above. \par
The action for type II models coupled to an external static gauge potential
${\cal{A}}_\mu=$$({\cal{A}}_0=0;{\bf{\cal{A}}}({\bf{x}}))$ can be obtained
by replacing ${\cal{D}}_\mu$ in (3.1) (and setting $J_0=0$) by
$$\hat{\cal{D}}_0={\cal{D}}_0-i\kappa{\cal{F}}_0\ ;\
\hat{\cal{D}}_i=D_i(A+{\cal{A}})-i\kappa F_i \eqno(3.3)$$
where ${\cal{F}}_0=\partial_1{\cal{A}}_2-\partial_2{\cal{A}}_1$ is the
external magnetic field.\par
At this level, some comments concerning the possible physical interpretation
of type II models are in order. The non minimal coupling defined by
(3.2) has been proposed and discussed in \cite{JSGW} as providing a possible
alternative description of composite anyonic objects. There, the usual minimal coupling
to a Chern-Simons statistical gauge potential controling the attachment
of an infinitesimaly thin flux tube to the charge carriers which basically
gives rise to standard (Chern-Simons) anyons is replaced by
the non minimal coupling (3.2) to a MCS statistical gauge potential with
$\kappa$ fixed to a specific value \cite{JSGW} to be given below. This alternative
way produces non standard anyons. In this respect, type II models can be
viewed as modeling in a second quantization framework the planar dynamics
of a system of charged non standard anyonic composite where now the statistical
interaction realizing the flux attachment has a finite range. This is
presented in more detail in the appendix for the sake of completeness.
Notice that (3.1) cannot be viewed as the naive non relativistic limit of a
model involving a relativistic scalar coupled through (3.2) to a MCS action.\par
Let us consider type II models in the presence of an uniform background
charge as described by (3.1)-(3.2). The equations of motion together with the gauge invariant matter current are
$$
-\frac{1}{e^2} \ \epsilon_{i j} \partial^i F^j + \eta F_0 + \kappa
 \epsilon_{i j} \partial^i j^j + j_0 - J_0 =0
 \eqno{(3.4a),}
$$
$$
\frac{1}{e^2} (\partial_i F_0 - \partial_0 F_i) + \eta \epsilon_{i j} F^j
-\kappa (\partial_i j_0-\partial_0 j_i) + \epsilon_{i j} j^j=0
\eqno{(3.4b),}
$$
$$
i {\cal D}_0 \phi + \frac{1}{2 m} \ {\cal D}_i {\cal D}_i \phi = 
 \frac{\delta V}{\delta \phi^\dag} \qquad\quad  ({\rm and \  h.c.} ) \ ,
\eqno{(3.4c),}
$$
$$ j_0=\phi^\dag \phi ; \ \ j_i =\frac{i}{2m} (\phi^\dag {\cal D}_i \phi-
({\cal D}_i \phi)^\dag \phi ) \ .
\eqno{(3.5a;b).}
$$
Now, when $\kappa=-1/(\eta e^2)$ which we assume from now on and upon setting
$$ G_0 = F_0 + \frac{1}{\eta} j_0  ;
\ \ G_i=F_i + \frac{1}{\eta} j_i
\eqno{(3.6a;b),} $$
eqn.(3.4a;b) can be rewritten as
$$
-\frac{1}{e^2} \epsilon_{ij} \partial^i G^j + \eta G_0 - J_0 =0
 ; \ \ \frac{1}{e^2} (\partial_i G_0-\partial_0 G_i) + 
 \eta \epsilon_{ij}  G^j=0
\eqno{(3.7a;b)}
$$
and are solved by $G_0=J_0/\eta$ and $G_i=0${\footnote{ Note
that, when $J_0=0$, (3.7a;b) are
solved by $F_i =-\frac{j_i}{\theta},F_0 =-\frac{j_o}{\theta}. $
This latter relation is
nothing but the usual anyonic relation ( Gauss law constraint) enforcing
the proportionality between the magnetic field and matter density that
would be obtained also within pure Chern-Simons theory with minimal
coupling to scalars}}. Therefore in the static regime, the relevant equations
take the form 
$$
-\frac{a^\prime}{r} + \frac{1}{\eta}(f^2-J_0)=0 \ ;
-\Omega{a_0^\prime} + \frac{f^2}{m
\eta} \ \frac{a}{r} = 0
\eqno{(3.8a,b),}
$$
$$
\frac{1}{2m} \Delta f+f\left(a_0+\frac{1}{\eta e^2}{{a^\prime}\over{r}} 
-\frac{1}{2m}(\frac{a}{r}+\frac{1}{\eta e^2}a_0^\prime)^2 \right) = \frac{\delta V}{\delta\phi^\dag}\vert_{\phi^\dag=f}
\eqno{(3.8c),}
$$
where we have used (2.4) and we have defined $\Omega=(1-{{f^2}\over{m\eta^2 e^2}})$.
The conjugate momenta for the fields are 
$$
\Pi_{\phi^+}=\Pi_{A_0}=0 ; \ \Pi_{\phi}=i \phi^+ ; \ \Pi_{A_i} =\frac{1}{e^2}
F_{0i}-\frac{\eta}{2} \epsilon_{ij} A^j+ \frac{1}{\eta e^2} \epsilon_{ij} j^j 
\ (i=1,2)
\eqno{(3.9a;b;c)}
$$
from which one obtains the Hamiltonian given by
$$
{\cal{H}}_2 = \int d{\bf x} \left[ \frac{1}{2 e^2} F_0^2
 + \frac{1}{2 e^2}\Omega {\bf{F}}^2 +
  \frac{1}{\eta e^2} F_0 \phi^\dag \phi + \frac{1}{2m} \vert {\bf{D}}\phi \vert^2 +
  V(\phi) \right]
\eqno{(3.10)}
$$
where $D_i =\partial_i - i  A_i $. The
positivity of (3.10) has been discussed in \cite{LATSOR}(see also
\cite{VORTMCSREL}) . 
Here, we will assume that
$\phi^\dag \phi \leq m \eta^2 e^2 $ as in \cite{VORTMCSREL},
a condition which will have to be verified {\it{a posteriori}}
by any vortex configuration. \par
The study of the possible existence of non trivial vortex configurations now
follows a way similar to the one described for type I models. Again, we restrict
ourselves to configurations saturating the lower bound for the energy. We
find that some type II models involving a local non polynomial potential do admit
vortex type solutions of the form (2.4) which minimize the energy. These
configurations are not present in type II models with polynomial potentials
(or they do not saturate the lower bound for the corresponding energy).\par
To see that, one first observes that the static energy functional can be
cast into the form
$$H_2=\tilde{H_{20}}+\int d{\bf{x}}{{1}\over{2m}}(f^\prime\pm
{{fa}\over{r{\sqrt{\Omega}}}})^2\mp\eta^2
e^2{\sqrt{\Omega}}{{a^\prime}\over{r}}+[V(f)-{{1}\over{2e^2\eta^2}}(f^4-J_0^2)]
\eqno(3.11)$$
where $\tilde{H_{20}}=\pm2\pi \eta^2e^2[a{\sqrt{\Omega}}]_0^\infty$ collects boundary
contributions and we have used (3.8a;b). By further making use of (3.8a),
(3.11) can be conveniently reexpressed as
$$H_2=H_{20}+H_2^\prime+H_{2V} \eqno(3.12a),$$
$$H_{20}=\tilde{H_{20}}+{{J_0}\over{\eta e^2}}\Phi\ ; \ H_2^\prime=\int d{\bf{x}}{{1}\over{2m}}(f^\prime\pm
{{fa}\over{r{\sqrt{\Omega}}}})^2 \eqno(3.12b;c),$$
$$H_{2V}=\int d{\bf{x}}[V(f)-{{1}\over{2e^2\eta^2}}(f^2-J_0)^2\mp
\eta e^2(f^2-J_0)(1-{{f^2}\over{m\eta^2 e^2}})^{{1}\over{2}}] \eqno(3.12d), $$
so that if the potential is choosen to take the following non polynomial form
$$V(\phi)={{1}\over{2e^2\eta^2}}((\phi^\dag\phi)-J_0)^2\pm\eta
e^2(\phi^\dag\phi-J_0)(1-{{\phi^\dag\phi}\over{m\eta^2 e^2}})^{{1}\over{2}}
\eqno(3.13),$$
which will be commented in a while, $H_2$ is minimized by configurations verifying
$$f^\prime\pm
{{fa}\over{r{\sqrt{\Omega}}}}=0 \eqno(3.14).$$
By combining (3.8a-c) with (3.14), one obtains  
$$\Delta f-{{f^{\prime
2}}\over{f{{\Omega}}}}=\mp{{f(f^2-J_0)}\over{\eta{\sqrt{\Omega}}}}
\eqno(3.15)$$
and
$$a_0=\pm\eta e^2{\sqrt{\Omega}} \eqno(3.16)$$
indicating that the corresponding configurations have a non vanishing
electric field. Note that we have determined the constant term in (3.16)
stemming from the integration of (3.8b) by requiring 
that it cancels all the terms linear in $f$ appearing in (3.8c).\par
Equation (3.15) has some interesting limits. For small matter density,
$f^2<<me^2\eta^2$ and $\Omega$ can be approximated by 1. Then, (3.15)
reduces to a non linear elliptic equation of the type (2.13) which has
physical solutions $f^2\to J_0$ for $r\to\infty$ ($f^2\to 0$ for $r\to 0$)
provided the upper (resp. lower) sign is choosen when $\eta<0$ (resp.
$\eta>0$). Note that positivity requires $J_0<<me^2\eta^2$ (since one must
have $\phi^\dag\phi<<me^2\eta^2$ in this regime). These solutions provide therefore
approximate solutions for (3.15)
in a small density regime. The corresponding configurations still carry a
charge proportional to the flux. When $J_0=0$, (3.15) simply reduces to a
Liouville equation.\par
By using (2.4d;e), the expression for $H_{20}$ in (3.12) becomes
$$H_{20}=2\pi n(\eta e^2)(\pm\eta^3e^2-J_0) \eqno(3.17)$$
and is proportional to the magnetic flux. It represents the positive lower bound 
for the static energy provided $n<0$
(resp. $n>0$) for $\eta>0$ (resp. $\eta<0$) corresponding to a positive
(resp. negative) magnetic flux.\par
The non linear potential (3.13) is depicted on the figure 1 for different
values of the ratio $J_0/(m\eta^2e^2)$ (and for $\eta>0$ and
$m/\eta^2e^2=1${\footnote{Other choices for theses parameters do not change
significantly the behaviour of the potential}}, the lowest (solid-line) curve
corresponding to $J_0=0$. It exhibits a symmetry
breaking shape with a minimum obtained at some
$\vert\phi_0\vert\le m\eta^2e^2$ which coincides with $J_0$,
$\vert\phi_0\vert=J_0$, when $J_0=m\eta^2e^2$. This potential has a somehow 
unusual expression in
that the usual "symmetry breaking" term (1st term in (3.13)) is supplemented
by an additional non linear term. Its origin can be traced back by first
noticing that non minimal electric field dependant couplings appearing in
(3.2) generate an extra contribution to the (electric) energy (see second
term in (3.10)). This gives rise to a (non linear) term in the static energy
which, if one insists on obtaining non trivial vortex solutions (i.e.
solutions with non constant matter density) that saturate the lower bound
for the energy, cannot be compensated by a polynomial term of finite degree
in $\phi^\dag\phi$. Correspondingly, in type II models with polynomial
potentials, the minimum for the energy is reached only by those
configurations having constant matter density (and/or vanishing gauge
potential). Vortex solutions with non constant matter density
possibly occur but they do not corresponds to a minimum for the energy.\par 
We have solved numerically eqn. (3.15). The resulting behaviour for the
matter density is depicted on fig.2 for different values of $J_0/m\eta^2e^2$
(with $\eta>0$ and $m/\eta e^2=1$). We find numerically that physical
configurations have the following asymptotic behaviour
$$f^2\sim r^2\ ,r\to 0\ ;\ f^2\to J_0\ ,r\to\infty  \eqno(3.18),$$
provided $J_0\le m\eta^2e^2$. The corresponding behaviour for the magnetic
field can be obtained from (3.8a). The magnetic field reaches its maximum
$F_0=J_0/\eta$ at the origin, decreases smoothly as $r$ increases and
vanishes at the infinity. Roughly speaking, matter is repelled away from the
area where the magnetic field is concentrated. It can be verified
numerically that the closest $J_0$ stands to the value $m\eta^2e^2$, the
fastest $f^2$ approaches its asymptotic plateau $f^2\sim J_0$ so that when
$J_0=m\eta^2e^2$, the solution of (3.15) is obtained only for constant
matter density $f^2=J_0$.\par
The above analysis can be extended to the case where type II models are
coupled to an external static gauge potential ${\cal{A}}_\mu$
$({\cal{A}}_0=0;{\bf{{\cal{A}}}}({\bf{x}}))$ instead of an uniform
background charge density. The corresponding action together with the
equations of motion are obtained by substituting (3.2) by (3.3) in (3.1) and
(3.4), (3.5) and further setting $J_0=0$ and $\kappa=-1/\eta e^2$, this
latter constraint insuring that (3.7) and (3.6) still hold (with (3.2)
replaced by (3.3)). Notice that the action coincides with (A.9) (in which
${\cal{A}}_0=0$ and $\gamma=-1/(2\eta)$. We consider here the case where the
vector potential ${\bf{\cal{A}}}$ gives rise to an external magnetic field
localized at the origin,
${\bf{\cal{A}}}_i={{\alpha}\over{2\pi}}\epsilon_{ij}{{x^j}\over{\vert{\bf{x}}\vert^2}}$
(${\bf{x}}=(x_1,x_2)$ with $\alpha>0$. The relevant static equations of
motion are then deduced from (3.8) by replacing $a$ by $(a+\alpha)$ in
(3.8b), (3.8c) while (3.8a) (where now $J_0=0$) is unchanged. The
Hamiltonian for the system is
$${\cal{H}}({\cal{A}})=\int d{\bf x} \left[ \frac{1}{2 e^2} F_0^2
 + \frac{1}{2 e^2}\Omega {\bf{F}}^2 +
  \frac{1}{\eta e^2}(F_0+{\cal{F}}_0) \phi^\dag \phi + \frac{1}{2m} \vert
  {\bf{D}}(A+{\cal{A}})\phi \vert^2 +
  V(\phi) \right] \eqno(3.19)$$
from which we obtain after some algebraic calculation the static energy
functional
$$H({\cal{A}})={\hat{H}}_0+{\hat{H}}^\prime+{\hat{H}}_V\ ;\ 
{\hat{H}}_0=\pm2\pi\eta^2e^2[(a+\alpha){\sqrt{\Omega}}]_0^\infty+
{{\alpha}\over{\eta e^2}}f^2(0)  \eqno(3.20a;b)$$
$${\hat{H}}^\prime=\int d{\bf{x}}{{1}\over{2m}}(f^\prime\pm
{{f(a+\alpha)}\over{r{\sqrt{\Omega}}}})^2\ ;\ {\hat{H}}_V= 
\int d{\bf{x}}[V(f)-{{1}\over{2e^2\eta^2}}f^4\mp
\eta e^2f^2(1-{{f^2}\over{m\eta^2 e^2}})^{{1}\over{2}}] 
\eqno(3.20c;d).$$
Therefore, if one chooses again the non linear potential (3.13) (in which
$J_0=0$), the static energy is minimized by configurations such that
$f^\prime\pm
{{f(a+\alpha)}\over{r{\sqrt{\Omega}}}}=0$ from which one easily realizes
that the matter density still verifies the differential equation (3.15) (for
$J_0=0$) while the electric potential is still given by (3.16).\par
The latter differential equation reduces to a Liouville equation when
$f^2<<m\eta^2e^2$ (small density regime). In this limit, physically
admissible solutions are obtained when $\eta>0$ (resp. $\eta<0$) and the
upper (resp. lower) sign in (3.20) is choosen. One then obtains
${\hat{H}}_0=\pm\eta^2e^2(2\pi n)$ which represents the lower bound for the
energy provided $n>0$ (resp. $n<0$) for $\eta>0$ (resp. $\eta<0$). As for
the first situation studied in the first part of this section, the present
type II models with polynomial potentials 
do not support non trivial vortex solutions minimizing the
energy. In particular, for $V=\lambda(\phi^\dag\phi-v)^2$, a simple
calculation shows that the minimum for the energy is obtained for
configurations such that $f^2=v$, $A_0=0$ and $A_i+{\cal{A}}_i=0$.\par

\vskip 0,3 true cm
\section{Discussion and conclusion}
\vskip 0,3 true cm
We have studied the possible occurence of radially symmetric static vortex
configurations saturating the lower bound for the energy in two types of MCS
models which differ from each other by their gauge-invariant coupling 
to a non relativistic
massive scalar field. Both models are further coupled to an external uniform
charge background. In type I models, where the scalar has a minimal and
magnetic dipole coupling to the MCS gauge potential $A_\mu$, non trivial
vortex configurations
satisfying the above requirement do occur when a relation between the
strength of the magnetic coupling, the scalar and the $A_\mu$ masses is
satisfied. The relevant scalar potential must be $V(\phi)={{1}\over{2\eta^2
e^2}}(\phi^\dag\phi-J_0)^2$. The corresponding vortex configurations have a
zero electric field, carry
a (quantized) magnetic flux proportional to the charge and saturate the
lower bound for the energy which is proportional to the flux.\par
Type II models {\footnote{In these models, non trivial vortex configurations
must have a non identically zero electric field, a fact which is already
apparent in the equations of motion which in particular imply that
either the matter density and/or the gauge potential function $a(r)$ is zero
whenever the electric field is zero}} involve both magnetic dipole and electric-field dependant
couplings to the non relativistic scalar as described by (3.2) in addition
to the usual minimal coupling. In the present work, we have
assumed that the strength for the magnetic coupling reaches the special
value already considered in \cite{JSGW}. These models are related to 
the planar dynamics of non standards anyonic composite objects, as indicated in the appendix.
Type II models with polynomial potential
(of finite degree) cannot support non trivial vortex minimizing the energy.
This is due to the contributions coming from the electric field dependant
coupling terms generating a (non linear) term in the static energy (see
(3.11)) which cannot be compensated by a finite number of polynomial
potential terms. When the potential is polynomial, the minimum for the
energy is obtained for configurations with zero or constant matter density
and/or vanishing gauge potential. However, non trivial vortex configurations
appear within type II models involving a non polynomial potential whose
expression is given by (3.13). These vortex solutions have a non zero
electric field and still carry a charge proportional to the magnetic flux.
The differential equation (3.15) constraining the 
matter density $f^2$ reduces in the small matter density regime
($f^2<<m\eta^2e^2$) to a non linear elliptic equation somehow similar to the
one constraining the matter density for the vortex configurations obtained
within type I models. We have solved (3.15) numerically and found that
$f^2$ vanished at the origin and increases smoothly until it reaches some
asymptotic plateau whose value is fixed by the magnitude of the uniform
background charge. Finally, we have considered the case of type II models
coupled to an external gauge potential corresponding to a magnetic field
localized at the origin. Again, those type II models with the non linear
potential (3.13) support non trivial vortex solutions minimizing the
energy.\par

\vskip 1 true cm
{\bf{Acknowledgments}}: We are grateful to A. Comtet, P. Horvathy and J.
Stern for discussions and comments at various stages of the present work.
\pagebreak
\vskip 2 true cm

{\bf{APPENDIX}}
\vskip 0,3 true cm
Consider the usual Hamiltonian for a system of $N$ (quasi)particles moving
in a plane and submitted for instance to an external static gauge potential
${\cal{A}}_\mu$
$$H_{qp}={{1}\over{2m}}\sum_{I=0}^N(i{\bf{\partial}}_i^{(I)}+{\bf{\cal{A}}}_i({\bf{x}}_I))^2
+\sum_{I=0}^N{\cal{A}}_0({\bf{x}})+... \eqno(A.1)$$
where $m$ is the mass for the (quasi)particles, capital Latin indices
$I,J,...$ refer to the particles $I,J,...$,
${\bf{x}}_{IJ}$$={\bf{x}}_I-{\bf{x}}_J$, the upper indice appearing in the
derivative operator means that it acts on the $I$-th particle and the
ellipses corresponds to possible potential terms whose explicit form will
not influence significantly the present discussion. The second quantized
Lagrangian counterpart of (A.1) is readily found to be
$${\cal{L}}=i\phi^\dag(\partial_0-i{\cal{A}}_0({\bf{x}}))\phi+{{1}\over{2m}}\phi^\dag
(\partial_i-i{\cal{A}}_i({\bf{x}}))^2\phi+... \eqno(A.2),$$
where $\phi$=$\phi(t,{\bf{x}})$ is a complex scalar field. Following the usual
way to introduce a statistical gauge potential \cite{ANREV}, \cite{VORTREV}, we first define the
singular gauge transformation
$\Psi^\prime({\bf{x}}_1,...,{\bf{x}}_N)$$=\exp(i{{\gamma}\over{\pi}}
\sum_{I<J}\alpha_{IJ})\Psi({\bf{x}}_1,...,{\bf{x}}_N)$ acting on the $N$-particles wave
function where $\gamma$ is a real constant and $\alpha_{IJ}$ denotes the
angle between ${\bf{x}}_{IJ}={\bf{x}}_I-{\bf{x}}_J$ and, says the
$x$-axis{\footnote{Interchange of any two particles gives
$\alpha_{IJ}\to$$\alpha_{IJ}\pm\pi$ so that it changes the phase of the wave
function by $\exp(i\gamma)$ whose statistics is unchanged (resp. changed)
for $\gamma=k2\pi$ (resp. $\gamma=(2k+1)\pi$), $k\in \mathbb{Z}$.}}. The
Hamiltonian relevant to $\Psi^\prime$ can then be expressed as
$$H^\prime_{qp}={{1}\over{2m}}\sum_{I=0}^N(i{\bf{\partial}}_i^{(I)}+{\bf{\cal{A}}}_i({\bf{x}}_I)
+A_i({\bf{x}}_I))^2
+\sum_{I=0}^N{\cal{A}}_0({\bf{x}})+... \eqno(A.3)$$
where the statistical gauge potential carrying the Aharonov-Bohm type
singularities is given by
$$A_i({\bf{x}}_I)={{\gamma}\over{\pi}}\sum_{I\ne
J}\partial_i^{(I)}\alpha_{IJ}=
-{{\gamma}\over{\pi}}\sum_{I\ne
J}\epsilon_{ij}{{(x_I-x_J)^i}\over{\vert {\bf{x}}_I-{\bf{x}}_J\vert^2}}
\eqno(A.4).$$
This, translated into a second quantized formalism, yields
$${\cal{H}}_{qp}=\int d{\bf{x}}\big(-{{1}\over{2m}}\phi^\dag({\bf{x}})
(\partial_i-i{\cal{A}}_i({\bf{x}})-iA_i({\bf{x}}))^2\phi({\bf{x}})\big)+...
\eqno(A.5a)$$
and
$$A_i({\bf{x}})=-{{\gamma}\over{\pi}}\int d{\bf{y}}
\epsilon_{ij}{{(x^i-y^i)}\over{\vert{\bf{x}}-{\bf{y}}\vert^2}}\rho({\bf{y}}) \eqno(A.5b)$$
with
$\rho({\bf{x}})=\phi^\dag\phi({\bf{x}})$. The final step amounts to treat
the statistical gauge potential as a dynamical variable of some action which
must be suitably choosen and coupled to the matter part such that (3.5b) is
solution of the corresponding equations of motion. Two different
inequivalent ways do exist to achieve this goal. \par
The first most currently used possibility which gives rise to the standard
anyons \cite{ANREV} is obtained when the statistical gauge potential is involved in a
Chern-Simons action minimally coupled to the matter. This can be summarized
as follows: One notices {\footnote{Owing to $\partial^2\ln\vert{\bf{x}}-
{\bf{y}}\vert=2\pi\delta({\bf{x}}-{\bf{y}})$.}} that (A.5b) yields
$\epsilon^{ij}\partial_iA_j({\bf{x}})=2\gamma\rho({\bf{x}})$ (i). Then, by
further allowing a time dependance in $A$ and $\rho$, differentiating (i)
with respect to time and restoring Lorentz covariance through the
introduction of a scalar potential $A_0$, one obtains
$\epsilon^{ij}\partial_i(\partial_0A_j-\partial_jA_0)$$=2\gamma\partial_0\rho$
$=-2\gamma\partial_i{\cal{J}}^i$ where ${\cal{J}}_\mu$$=(\rho;{\cal{J}}_i)$
is the gauge invariant matter current and current conservation has been
used. This produces $\epsilon^{ij}(\partial_0A_j-\partial_jA_0)$
$=-2\gamma{\cal{J}}^i$ (ii). One then
easily realizes that (i) and (ii) can be obtained as the equations of motion
for
$S_{CS}=\int d^3x
\big(-{{1}\over{4\gamma}}\epsilon_{\mu\nu\rho}A^\mu\partial^\nu
A^\rho+A_\mu{\cal{J}}^\mu\big).$
Then, allowing $\phi$ to depend on time and restoring again the Lorentz covariance
through the introduction of $A_0$, one easily obtains the second quantized
Lagrangian version describing the planar dynamics of (quasi)particles with
Chern-Simons statistical interaction which is related to the standard anyons
\cite{ANREV}, \cite{VORTREV}
$${\cal{L}}=i\phi^\dag(\partial_0-i({\cal{A}}_0+A_0))\phi+{{1}\over{2m}}
\phi^\dag(\partial_i-i({\cal{A}}_i+A_i))^2\phi
-{{1}\over{4\gamma}}\epsilon_{\mu\nu\rho}A^\mu\partial^\nu A^\rho+...
\eqno(A.6)$$
where still ${\cal{A}}_\mu={\cal{A}}_\mu({\bf{x}})$ and corresponds
therefore to (A.2) with minimal coupling to a Chern-Simons action for $A_\mu$.\par
The alternative possibility has been proposed and further discussed in
\cite{JSGW}. It is obtained by noticing that (A.5b) is also solution of the equations
of motion stemming from a MCS action for the statistical field with minimal
and non minimal coupling to matter as given in (3.2), provided the strength
$\kappa$ of the non minimal coupling is fixed to a special value. To see
that, consider the following action for the statistical gauge potential
$$S=\int d^3x
-{{1}\over{4e^2}}F_{\mu\nu}F^{\mu\nu}-{{1}\over{8\gamma}}\epsilon_{\mu\nu\rho}A^\mu
 F^{\nu\rho}+A_\mu{\cal{J}}^\mu+\kappa F_\mu{\cal{J}}^\mu \eqno(A.7)$$
where $F_\mu$ is the dual field strength. When
$\kappa={{\gamma}\over{e^2}}$, the corresponding equation of motion can be
expressed as
$$-{{1}\over{e^2}}\epsilon_{\alpha\nu\rho}\partial^\mu(F^\rho-2\gamma{\cal{J}}^\rho)
-{{1}\over{2\gamma}}(F_\alpha-2\gamma{\cal{J}}_\alpha)=0 \eqno(A.8)$$
which is formally similar to the equations of motion for a free MCS theory
and are solved by $F_\mu=2\gamma{\cal{J}}_\mu$ whose time and space
components are nothing but equations (i) and (ii) given above. Accordingly,
the resulting second quantized Lagrangian version obtained from (A.5) can
be found to be given by
$${\cal{L}}_{II}=i\phi^\dag(\partial_0-i({\cal{A}}_0+A_0))\phi+
{{2\gamma}\over{e^2}}F_0\phi^\dag\phi$$
$$+{{1}\over{2m}}
\phi^\dag(\partial_i-i({\cal{A}}_i+A_i)-i{{2\gamma}\over{e^2}}F_i)^2\phi
-{{1}\over{4e^2}}F_{\mu\nu}F^{\mu\nu}-{{1}\over{8\gamma}}\epsilon_{\mu\nu\rho}A^\mu
 F^{\nu\rho}+... \eqno(A.9)$$
where $F_0=\epsilon_{ij}\partial^iA^j$ and $F_i=\partial_iA_0-\partial_0A_i$
which is similar to the Lagrangian defining type II MCS models considered in
section 3 (with $\gamma=-{{1}\over{2\eta}}$).\par
We note that the Lagrangian (A.6) is an important piece appearing in the
Landau-Ginzburg type effective model proposed in \cite{KLZ} to describe some of the
global properties of the Quantum Hall Effect. In this latter description, 
one of the building ingredient was the reformulation
  of the problem of interacting fermions in an external magnetic field as a
  problem of interacting bosons with (minimal) coupling to a Chern-Simons
  gauge field, the statistical field,  controlling the statistical
  transmutation of bosons to fermions. Now, in the alternative description
of anyons proposed in \cite{JSGW}, the statistical transmutation is obtained through the minimal
and suitable non-minimal coupling of a MCS statistical gauge field to matter.
In this spirit, (A.9) can be viewed as a modification of the above effective theory in which 
statistical transmutation is controlled now by the coupling of a MCS
statistical field with a suitable coupling to the scalar.
\pagebreak

\clearpage

\begin{figure}
\begin{center}
\includegraphics{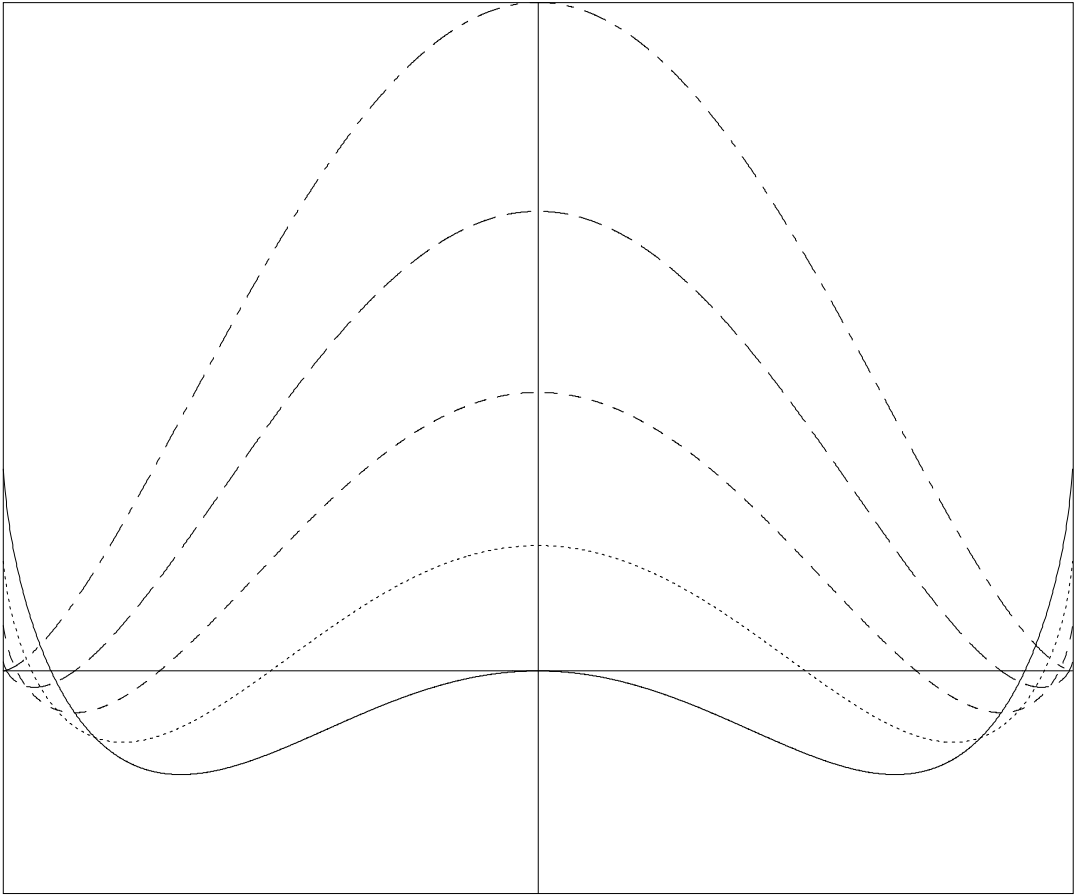}
\end{center}
\caption{Qualitative shape of the non-linear potential, plotted for
different values of the ratio $J_0/m\eta^2e^2$, assuming $\eta>0$ and
$m/\eta e^2=1$. Choosing $\eta<0$ and/or other values for $m/\vert\eta\vert 
e^2$ does not change significantly the behaviour of the potential. From the
lowest curve to the uppermost one, one has
$J_0/m\eta^2e^2 =0,1/4,1/2,3/4,1$.}
\end{figure}

\begin{figure}
\begin{center}
\includegraphics{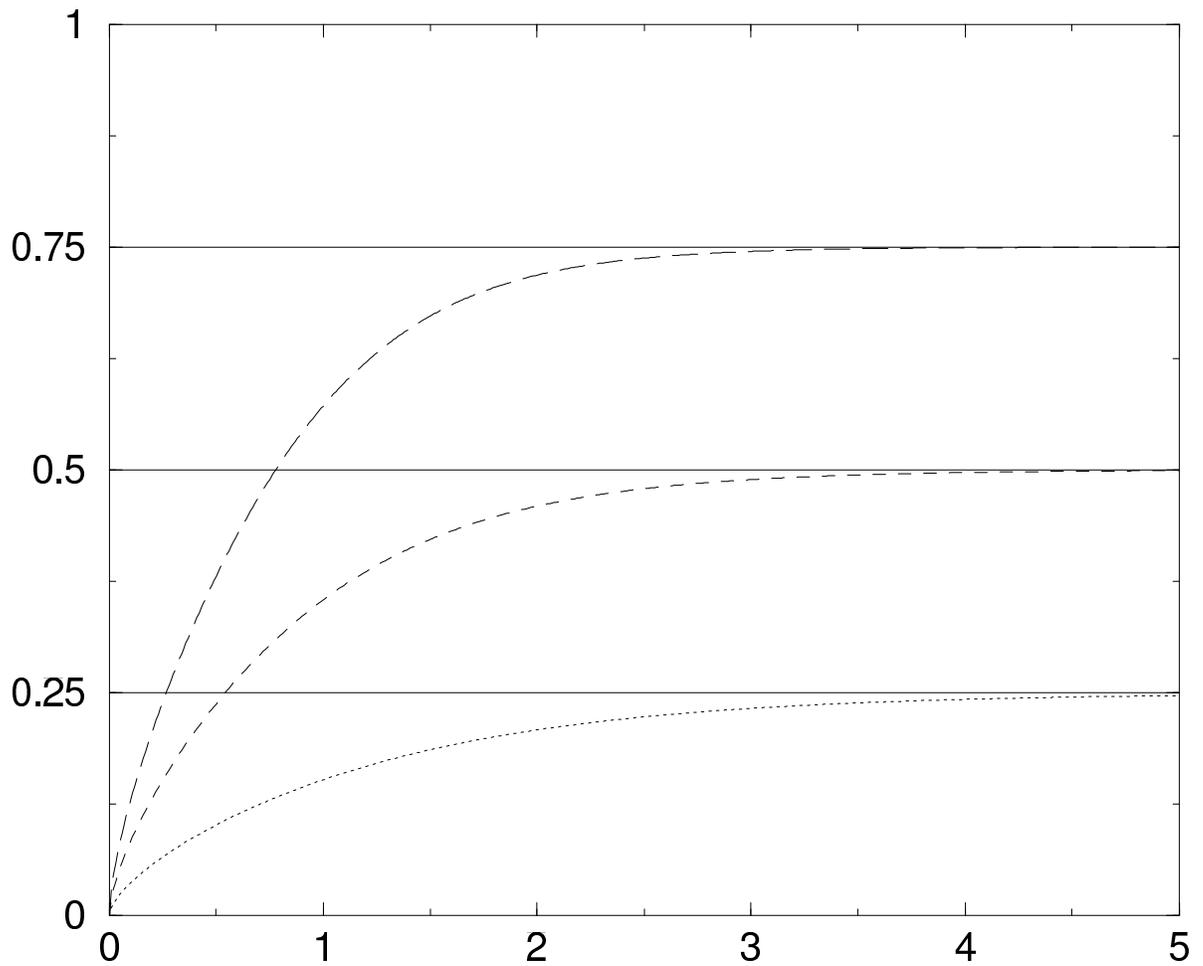}
\end{center}
\caption{Radial behaviour for the matter density. The quantity
${{f^2}\over{m\eta^2e^2}}$ (vertical axis) is plotted versus
$\rho\equiv{{r}\over{\eta e^2}}$ (horizontal axis). As on fig.1, $\eta>0$
and $m/\eta e^2=1$. The three curves from bottom to top correspond
respectively to ${{J_0}\over{m\eta^2e^2}}$ $=1/4,1/2,3/4$, indicated by the
three horizontal asymptotic lines.}
\end{figure}

\end{document}